\RequirePackage{fix-cm}
\documentclass[smallextended]{svjour3}       
\smartqed  
\usepackage{graphicx}
\begin{document}

\title{Elastic measurements of amorphous silicon films at mK temperatures
}


\author{Andrew Fefferman         \and
        Ana Maldonado \and
        Eddy Collin \and
        Xiao Liu \and
        Tom Metcalf \and
        Glenn Jernigan
}


\institute{A. Fefferman \and A. Maldonado \and E. Collin \at
              Institut N\'{e}el, CNRS and Universit\'{e} Grenoble Alpes, Grenoble, France \\
              \email{andrew.fefferman@neel.cnrs.fr}
           \and
           X. Liu \and T. Metcalf \and G. Jernigan \at
              Naval Research Laboratory, Washington, D.C., USA
}

\date{Received: date / Accepted: date}

\maketitle

\begin{abstract}
The low temperature properties of glass are distinct from those of crystals
due to the presence of poorly understood low-energy excitations. The
tunneling model proposes that these are atoms tunneling between nearby
equilibria, forming tunneling two level systems (TLSs). This model is rather successful,
but it does not explain the remarkably universal value of the mechanical
dissipation $Q^{-1}$ near 1 kelvin.
The only known exceptions to this universality are the $Q^{-1}$ of certain
thin films of amorphous silicon, carbon and germanium. Recently, it was
found that $Q^{-1}$ of amorphous silicon (a-Si) films can be reduced by two
orders of magnitude by increasing the temperature of the substrate during
deposition. According to the tunneling model, the reduction in $Q^{-1}$ at 1
kelvin implies a reduction in $P_{0}\gamma ^{2}$, where $P_{0}$ is the
density of TLSs and $\gamma $ is their coupling to phonons. In this
preliminary report, we demonstrate elastic measurements of a-Si films down
to 20 mK. This will allow us, in future work, to determine
whether $P_{0}$ or $\gamma $ is responsible for the reduction in $Q^{-1}$
with deposition temperature.
\end{abstract}

\section{Introduction}
At temperatures below a few kelvin, the thermal, mechanical and dielectric
properties of glass are dominated by low energy excitations (LEEs) that are
not present in crystals. The first sign of these LEEs was the discovery that
the heat capacity of glass below 1 kelvin is much greater than that of
crystals and has a nearly linear temperature dependence, compared with the
cubic temperature dependence observed in insulating crystals \cite{Zeller71}.
While the identity of the LEEs remains uncertain \cite{Agarwal13,Vural11}, the
tunneling model proposes that they are two level systems (TLSs) formed by
atoms tunneling between nearby equilibria in the disordered lattice of the
glass \cite{Anderson72,Phillips72}. The tunneling model successfully describes
many aspects of the low temperature behavior of glass, but one major
shortcoming is its failure to explain the remarkable universality of the
mechanical dissipation $Q^{-1}$ of glass near 1 kelvin. The level of the
$Q^{-1}$ plateau near 1 kelvin is universal within a factor of 20 centered at
$5\times10^{-4}$ and is insensitive to chemical composition, impurity
concentration, preparation of the samples including heat treatment,
measurement frequency over 9 orders of magnitude and stiffness of the glass
over four orders of magnitude \cite{Pohl02}. The only known exceptions to this
universality are in four-fold coordinated glasses:\ certain thin films of
amorphous silicon, carbon and germanium have $Q^{-1}$ up to a factor of 100
below the glassy range \cite{Pohl02}.

The origin of the reduced $Q^{-1}$ in
these materials was uncertain, but it was recently found that the $Q^{-1}$ of
a-Si films can be tuned by varying the temperature of the substrate during
deposition:\ increasing the substrate temperature during deposition from room
temperature to 400 Celsius results in a reduction of $Q^{-1}$ of about two
orders of magnitude \cite{Liu14}. The decrease in $Q^{-1}$ was explained in
terms of enhanced surface energetics at high growth temperatures \cite{Liu14}.
Indeed, the high mobility of atoms within a few nanometers of the glass
surface during deposition is known to produce highly stable glasses that are
low on their energy landscape \cite{Swallen07}. Despite the reduced $Q^{-1}$
of the a-Si films deposited at high temperatures, they remain fully amorphous,
as demonstrated by Raman spectroscopy, electron and x-ray diffraction and high
resolution cross section transmission electron microscopy \cite{Liu14}. In
addition to vapor deposition, the four-fold coordination of the a-Si films may
also play a role in suppressing TLSs in this material \cite{Phillips72,Liu14}.

In this preliminary report, we demonstrate low temperature elastic
measurements of an a-Si film deposited at room temperature. This technique will allow
us, in future work, to separately determine the density of TLSs and their
coupling to phonons. A central assumption of the tunneling model is that the
TLSs have a broad distribution of parameters with spectral density $P\left(
\Delta,\Delta_{0}\right)  =P_{0}/\Delta_{0}$, where $\Delta$ is the asymmetry
in the depth of the two wells of the TLS and $\Delta_{0}$ is the tunneling amplitude \cite{Phillips87}. The energy splitting of a TLS is
$E=\sqrt{\Delta^{2}+\Delta_{0}^{2}}$. Strain $\epsilon$ in the glass has
little effect on $\Delta_{0}$ but changes the asymmetry of the TLSs by
$\gamma\epsilon$, thus changing $E$. Consequently, after the glass is deformed, the fraction of TLSs in
their excited states no longer satisfies thermal equilibrium and the ensemble
of TLSs relaxes at a rate
\begin{equation}
\tau^{-1}=\frac{3\gamma^{2}}{v^{5}}\frac{E\Delta_{0}^{2}}{2\pi\rho\hbar^{4}%
}\coth\left(  E/2k_{B}T\right)%
\label{eq:tauinv}
\end{equation}
This relaxation causes softening of the glass as well as dissipation. If all the TLSs had the same energy, 
a maximum in dissipation would occur when the relaxation rate of
the TLS matches the driving frequency, $\omega=\tau^{-1}$. For the distribution of TLS parameters given above, the dissipation due to relaxation is 
temperature independent for $\omega<<\tau^{-1}$ and cubic in temperature for $\omega>>\tau^{-1}$. Furthermore, a phonon can be
absorbed by a subset of the TLSs that have energies matching the frequency of
the phonon. This causes significant dissipation when $\hbar\omega\ge k_{B}T$ (not the case in this work) as well as a contribution to the sound
speed that increases logarithmically with temperature.

Liu \emph{et al.} observed a plateau in $Q^{-1}$ of a-Si films between 300 mK and 10 K.  The films deposited at room
temperature had $Q^{-1}=2\times10^{-4}$ in this temperature range. According to the tunneling model,
the plateau occurs because $\omega<<\tau^{-1}$ for the subset of TLSs that dominate the mechanical response,
i.e., those with $E\approx\Delta_0\approx k_{B}T$.
The value of $Q^{-1}$ at this plateau is
$Q_{0}^{-1}=\pi P_{0}\gamma^{2}/2 \rho v^{2}$, where $\rho$ is the mass density of
the glass and $v$ is its sound speed \cite{Fefferman10}. As the temperature decreases, we expect a maximum in
the sound speed at the temperature where $\omega\approx\tau^{-1}$ . The maximum occurs
because the contribution of relaxation to the softening becomes negligible
compared with that of resonant phonon/TLS interactions at the lowest temperatures.

\section{Experimental Details}

\begin{figure}
  \includegraphics[width=12cm]{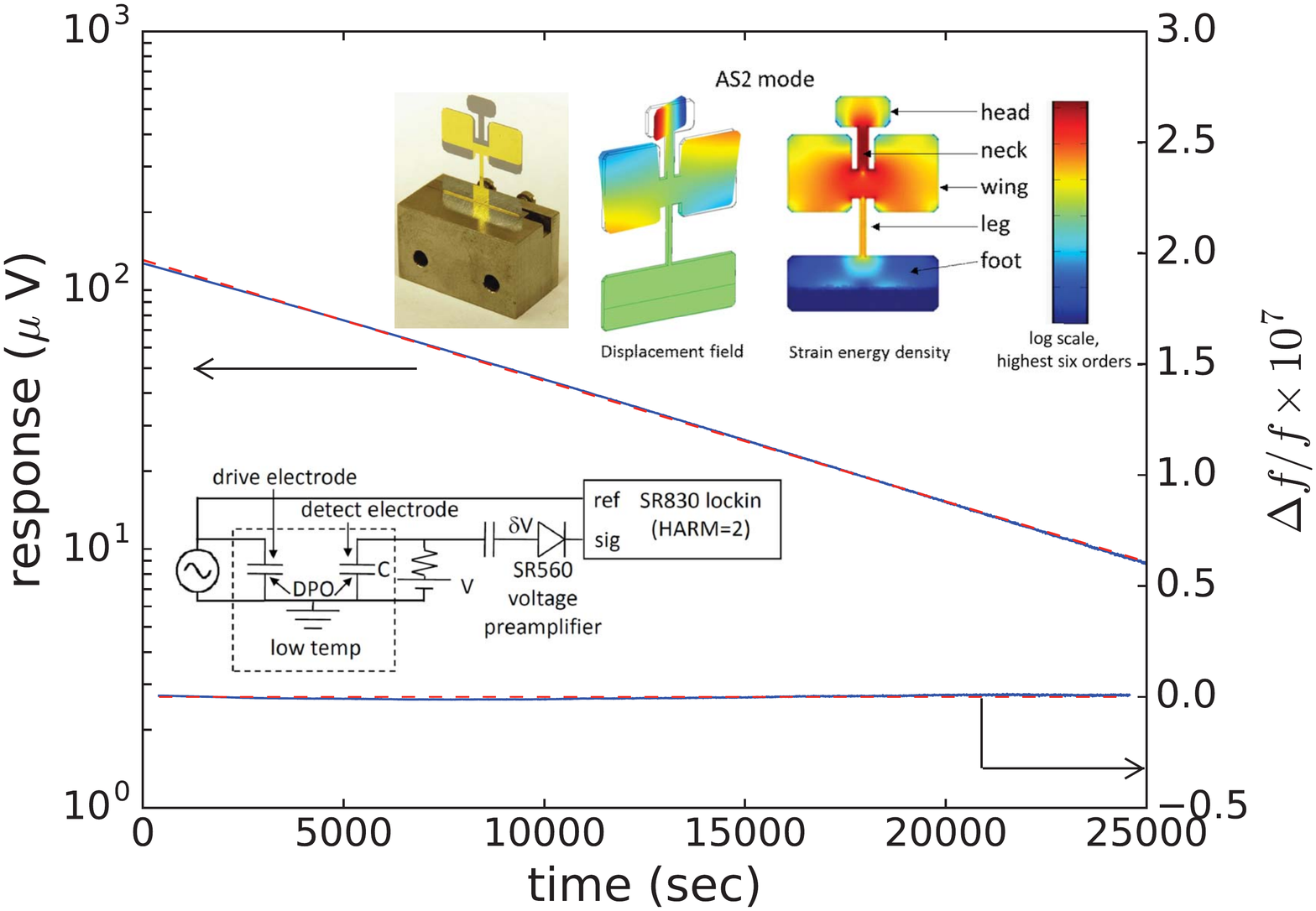}
\caption{Inset: Photograph of a silicon double paddle oscillator (DPO) and calculation of its motion. The gold film is used for electrostatic drive and detection of its motion and does not cover the neck, which is the sample region. Also shown is the circuit used for drive and detection. Main Panel: Measurement of a bare DPO (i.e. without a sample) at 10 mK. The free decay of its response (left axis) and the corresponding time evolution of its resonant frequency shift $\Delta f/f$ (right axis) are shown.  The horizontal dashed red line is a guide to the eye and the other dashed red line is a linear fit, yielding $Q^{-1}=6.0\times10^{-9}$. The variation of $\Delta f/f$ during the ringdown was negligible compared with the temperature dependence of this quantity (Fig. 2). (Color figure online.)}
\label{fig:1}
\end{figure}

The sound speed and $Q^{-1}$ of the a-Si film was measured using a double
paddle oscillator (DPO) like the one pictured in the inset of Fig.
\ref{fig:1}. The DPO has been characterized in detail in previous work \cite{Liu01,Spiel01,Fefferman10,Liu12}. It
is etched from a 300 $\mu$m thick wafer of crystalline silicon. Of the ten
lowest modes of vibration, the AS2 mode around 5.5 kHz has by far the lowest
$Q^{-1}$, which reaches below 10$^{-8}$ at the lowest temeratures. The other
modes have $Q^{-1}$ around 10$^{-6}$ at low temperatures. The exceptionally
low $Q^{-1}$ of the AS2 mode is due to the low strain amplitude at the
clamping position. In this mode, the strain is concentrated in the upper
torsion rod, called the neck, where the sample is deposited. A gold or
platinum metal film is deposited on the wings, leg and foot of the DPO to
facilitate electrostatic drive and detection of the motion and to thermalize
it. When the DPO is installed in its electrode structure, each wing is
separated from an electrode by a $\approx100$ $\mu$m gap, forming two of the
capacitors represented in the circuit in Fig. \ref{fig:1}. The DPO is clamped
in an Invar block, whose low thermal expansion coefficient minimizes the risk
of cracking the DPO while cooling. A calibrated carbon resistor was screwed to
the side of the Invar block for thermometry.

The resonant frequency of the DPO was measured using the ringdown technique with the circuit in Fig. \ref{fig:1}.
We first applied a driving voltage across the capacitor formed by the drive electrode and the
grounded metal film on the DPO. The driving voltage was at half the DPO resonant frequency since the force
on the DPO was proportional to the square of the drive voltage. To initiate the ringdown,
the driving voltage was reduced by a factor of 250 and its frequency was increased to 10 mHz above the DPO resonance,
so that the AS2 mode was effectively undriven.
On the detection side, a voltage source was used to bias the detection electrode at voltage $V$. A blocking capacitor
was added to protect the preamplifier from the bias voltage. Due to a 10 M$\Omega$ resistor in series with the voltage source and the
100 M$\Omega$ input impedance of the preamplifier, the charge on the detection electrode remained nearly constant on the
time scale of the DPO oscillations. Consequently, the fluctuations $\delta V$ of the voltage across the capacitor
$C$ formed by the detection electrode and the grounded DPO were
proportional to the capacitance fluctuations $\delta C$ caused by the motion of the
DPO: $\delta V/V=-\delta C/C$. The quantity $\delta V$ was measured using the ``Harm=2'' setting of an SR830 dual phase lockin amplifier. This setting causes
it to generate an internal reference signal at a frequency $f_{\mathrm{ref}}$ that is twice the
frequency of the signal input to its reference port, which was output by the function generator. The lockin output the phase $\theta$ of $\delta V$
relative to the internal reference signal. We determined the oscillation frequency of the DPO during the ringdown
(i.e. its resonant frequency $f_{\mathrm{res}}$) using $f_{\mathrm{res}}=f_{\mathrm{ref}}+(1/2\pi)d\theta/dt$.

\section{Results and Discussion}

The amplitude of $\delta V$ during the ringdown of a
bare DPO at 10 mK is plotted as a function of time in Fig. \ref{fig:1}. As
expected, it has a nearly perfect exponential time dependence, as shown by the
dashed red line. The corresponding time dependence of $f_{\mathrm{res}}$ is also
shown in the figure. The
horizontal red dashed line shows that the response of the
DPO is linear at these strain magnitudes and that fluctuations in
$f_{\mathrm{res}}$ are small compared with the temperature dependence of
$f_{\mathrm{res}}$, which will be discussed below.

\begin{figure}
  \includegraphics[width=12cm]{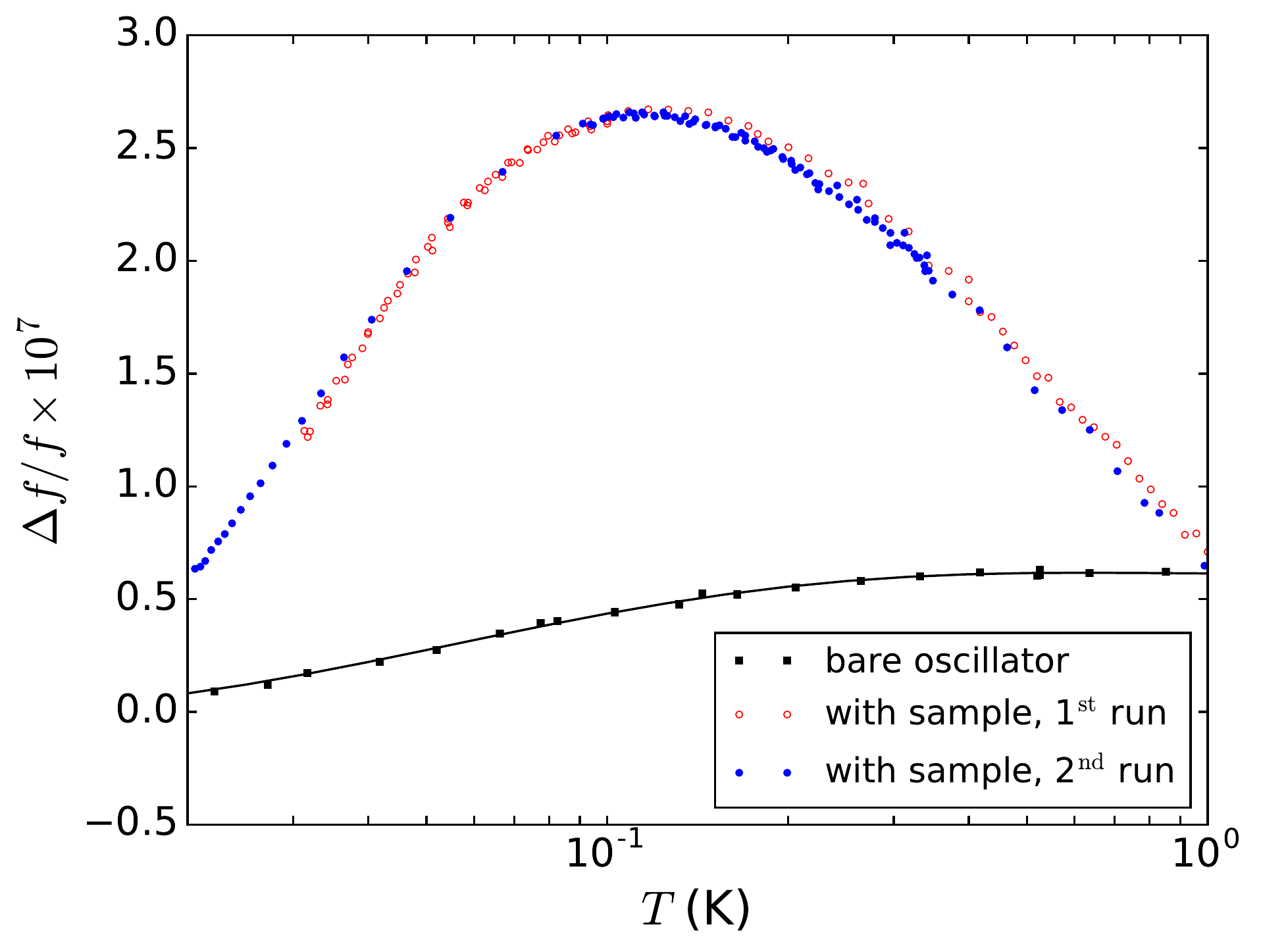}
\caption{Temperature dependence of the resonant frequency shift $\Delta f/f=(f-f_0)/f_0$ of a bare oscillator (called DPO1) [black] and the same DPO with a 336.1 nm thick amorphous silicon film deposited on its neck at 22 deg Celsius [blue and red]. The DPO was warmed to room temperature between the acquisition of the blue and the red data. The reference frequency $f_0$ has three different values for the three data colors.(Color figure online.)}
\label{fig:2}
\end{figure}

Figure \ref{fig:2} shows the temperature dependence of the resonant frequency
of the bare DPO obtained from measurements like those shown in Fig.
\ref{fig:1} as well as that of the DPO\ with an a-Si film. The coincidence of
the red and blue data points, which came from two different runs starting from
room temperature, demonstrates the reproducibility of the temperature dependence.

\begin{figure}
  \includegraphics[width=12cm]{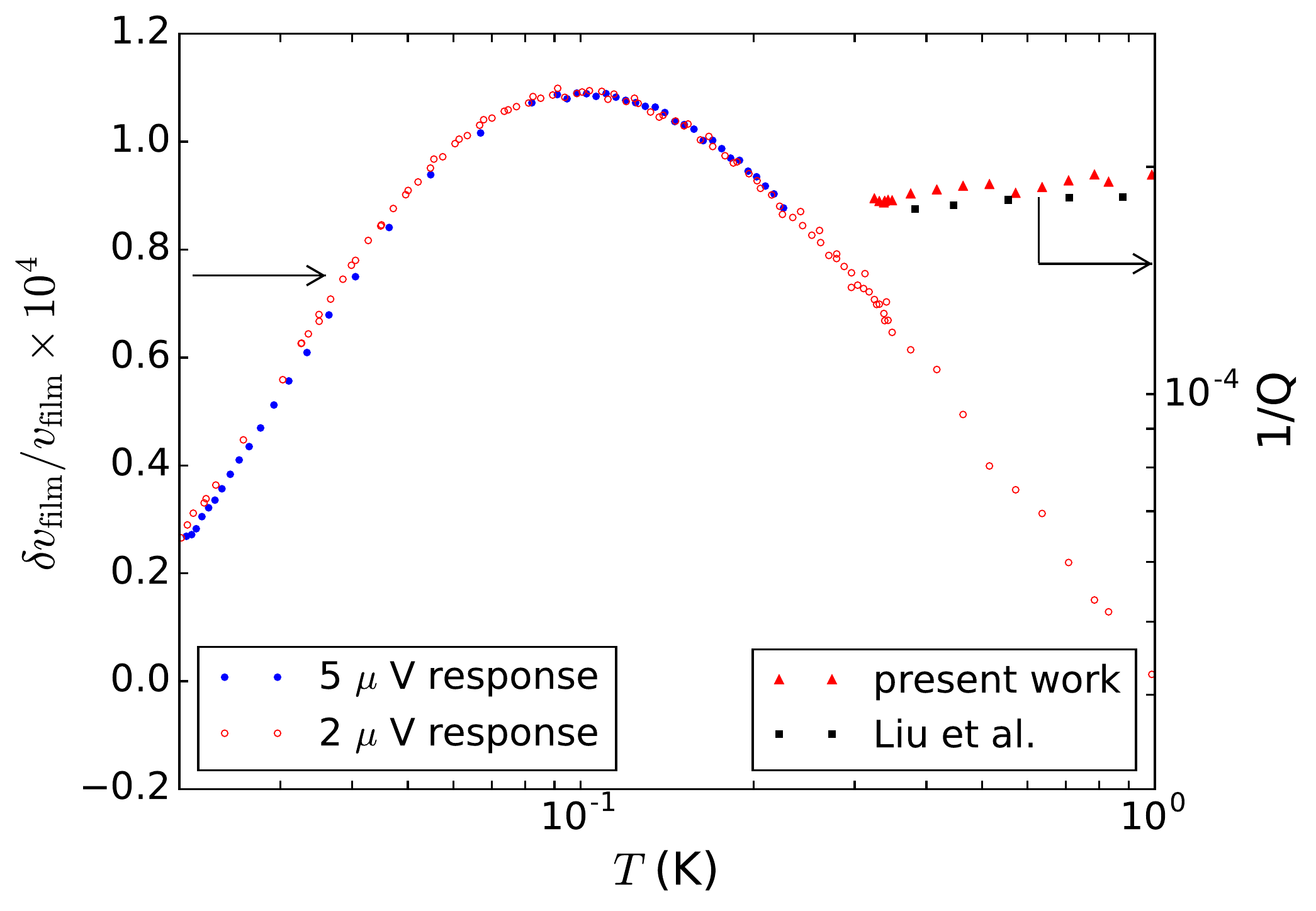}
\caption{Temperature dependence of the relative change in sound speed $\delta v_{\mathrm{film}}/v_{\mathrm{film}}=(G_{\mathrm{film}}-G_0)/2G_0$ in the amorphous silicon film, as determined from the measurements in Fig. 2 using Eq \ref{eq:Gfilm}. We set $G_0=37$ GPa. The lack of dependence of the sound speed on the response amplitude demonstrates that the measurements were made in the linear regime. Also shown is the agreement between $Q^{-1}$ 
of the present film and that measured in \cite{Liu14} (see text). (Color figure online.)}
\label{fig:3}
\end{figure}

In order to determine the temperature dependence of the sound speed in the film,
the contribution of the bare paddle was subtracted as follows. The resonant
frequency of the AS2 mode scales with the square root of the torsion constant
of the DPO neck $\kappa_{\mathrm{neck}}\propto wt^{3}$, where $w$ and $t$ are
respectively the width and thickness of the neck. Consequently, the frequency
shift upon depositing a sample on the DPO neck is \cite{Liu12}%
\begin{equation}
\frac{f_{\mathrm{osc}}-f_{\mathrm{sub}}}{f_{\mathrm{sub}}}=\frac{3t_{\mathrm{film}%
}G_{\mathrm{film}}}{t_{\mathrm{sub}}G_{\mathrm{sub}}}%
\label{eq:Gfilm}
\end{equation}
where $f_{\mathrm{osc}}$ and $f_{\mathrm{sub}}$ are respectively the frequencies
of the film-laden and bare DPO, and $G_{\mathrm{film}}$ and $G_{\mathrm{sub}}=62$
GPa are respectively the shear moduli of the film and the substrate. Since
$t_{\mathrm{film}}=336.1$ nm and $t_{\mathrm{sub}}=300\mu$m, this equation can be
used to calculate $G_{\mathrm{film}}$. The solid black line in Fig. \ref{fig:2}
is a polynomial fit to $f_{\mathrm{sub}}$ that was used for this purpose. The
resulting $\delta v_{\mathrm{film}}/v_{\mathrm{film}}=\delta G_{\mathrm{film}%
}/2G_{\mathrm{film}}$ is plotted in Fig. \ref{fig:3}. The blue and the red data
correspond to measurements at two different inital strain levels (each
ringdown measurement lasted 500 seconds), and the coincidence of these data
demonstrates that the measurements were made in the regime of linear response.
The intrinsic non-linear response due to TLSs was analysed theoretically in
\cite{Stockburger95}, and it will be studied experimentally in this material
in future work. Figure \ref{fig:3} also shows $Q^{-1}$ of the a-Si film studied here as well as 
$Q^{-1}$ of an a-Si film measured by Liu \emph{et al.} \cite{Liu14}, which was deposited at a slightly higher temperature (45 deg C). 
As expected, the $Q^{-1}$ of the film of Liu \emph{et al.} is slightly less than but almost the same as that of the present film. The superposition of these data demonstrate that our measurements of $Q^{-1}$ are consistent.

It has been proposed that, in a-Si, TLSs can only exist at imperfections in the four-fold coordinated atomic structure \cite{Phillips72,Liu14}.
This is in contrast to a-SiO$_2$, which has a more open structure where most atoms have extra degrees of freedom. Despite these structural differences, the temperature
dependence of the sound speed in a-Si below 1 kelvin (Fig. \ref{fig:3}) is qualitatively similar to
that of a-SiO$_2$ \cite{Fefferman08}. Indeed, in a-Si $\delta v_{\mathrm{film}}/v_{\mathrm{film}}$ has a
maximum at $T_{\mathrm{co}}=$100 mK and a logarithmic temperature dependence above and below
$T_{\mathrm{co}}$. From this result, we can make a rough estimate of $\gamma$, which could not be determined in the higher temperature
measurements of \cite{Liu14}. As explained above,
the contribution to the mechanical susceptibility from relaxation of TLSs is dominated by symmetric TLSs with energy
splittings approximately equal to the temperature of the glass \cite{Fefferman_thesis}. Setting $\Delta_{0}=E=T=T_{\mathrm{co}}=$100 mK
in Eq. \ref{eq:tauinv} and setting $\tau^{-1}(T_{\mathrm{co}})=\omega$ yields $\gamma\approx$ 1 eV,
compared with $\gamma=0.8$ eV in \cite{Fefferman08}. A more precise determination of $\gamma$ in a-Si requires a
fit of the tunneling model to both $\delta v_{\mathrm{film}}/v_{\mathrm{film}}$
and $Q_{\mathrm{film}}^{-1}$, which will be carried out in future work.

\begin{acknowledgements}
We acknowledge support from the ERC CoG grant ULT-NEMS No. 647917 and from the US Office of Naval Research.
\end{acknowledgements}


%
%

\end{document}